\documentclass[twocolumn,english,prl,showpacs,preprintnumbers,superscriptaddress]{revtex4-1}
\usepackage[utf8]{inputenc}
\usepackage{color}
\usepackage{amsmath}
\usepackage{amssymb}
\usepackage{graphicx}
\usepackage{textcomp}
\usepackage{dcolumn}
\usepackage{bm}
\usepackage[right]{eurosym}
\usepackage{float}
\usepackage[english]{babel}
\usepackage{blindtext}
\usepackage{babel}
\usepackage[normalem]{ulem}

\begin{document}

\title{Enhanced ionization of embedded clusters by Electron Transfer Mediated Decay in helium nanodroplets}

\author{A. C. LaForge}
\email{aaron.laforge@physik.uni-freiburg.de}
\affiliation{Physikalisches Institut, Universit{\"a}t Freiburg, 79104 Freiburg, Germany}
\author{V. Stumpf}
\author{K. Gokhberg}
\affiliation{Physikalisch-Chemisches Institut, Universit{\"a}t Heidelberg, 69120 Heidelberg, Germany}
\author{J. von Vangerow}
\author{F. Stienkemeier}
\affiliation{Physikalisches Institut, Universit{\"a}t Freiburg, 79104 Freiburg, Germany}
\author{N. V. Kryzhevoi}
\affiliation{Physikalisch-Chemisches Institut, Universit{\"a}t Heidelberg, 69120 Heidelberg, Germany}
\author{P. O'Keeffe}
\author{A. Ciavardini}
\affiliation{CNR - Istituto di Struttura della Materia, CP10, 00016 Monterotondo Scalo, Italy}
\author{S. R. Krishnan}
\affiliation{Department of Physics, Indian Institute of Technology - Madras, Chennai 600 036, India}
\author{M. Coreno}
\affiliation{CNR-Istituto di Struttura della Materia, CP10, 00016 Monterotondo Scalo, Italy}
\author{K. C. Prince}
\author{R. Richter}
\affiliation{Elettra-Sincrotrone Trieste, 34149 Basovizza, Trieste, Italy}
\author{R. Moshammer}
\author{T. Pfeifer}
\affiliation{Max-Planck-Institut f{\"u}r Kernphysik, 69117 Heidelberg, Germany}
\author{L. S. Cederbaum}
\affiliation{Physikalisch-Chemisches Institut, Universit{\"a}t Heidelberg, 69120 Heidelberg, Germany}
\author{M. Mudrich}
\affiliation{Physikalisches Institut, Universit{\"a}t Freiburg, 79104 Freiburg, Germany}

\begin{abstract}
We report the observation of electron transfer mediated decay (ETMD) involving magnesium (Mg) clusters embedded in helium (He) nanodroplets. ETMD is initiated by the ionization of He followed by removal of two electrons from the Mg clusters of which one is transferred to the He environment while the other electron is emitted into the continuum. The process is shown to be the dominant ionization mechanism for embedded clusters for photon energies above the ionization potential of He. For Mg clusters larger than 5 atoms we observe stable doubly-ionized clusters. Thus, ETMD provides an efficient pathway to the formation of doubly-ionized cold species in doped nanodroplets.   
\end{abstract}

\date{\today}

\maketitle

The interplay between electrons after single photon absorption has been a stimulating topic in atomic physics since its foundation. Specifically, processes such as shake-off in single photon double ionization~\cite{Pattard2003}, post collision interaction in Auger processes~\cite{Russek1986}, and autoionization of doubly-excited states~\cite{Madden1963} have been a fertile ground for studying electron correlation. Systems consisting of many weakly-interacting atoms or molecules additionally offer a unique environment for studying said correlation where new decay mechanisms, unavailable in atomic systems, become accessible between the constituents. In particular, Cederbaum and coworkers~\cite{Cederbaum1997} theoretically predicted a new decay mechanism, known as interatomic Coulombic decay (ICD), available in electronically excited weakly bound systems. In the case where local electronic decay is energetically forbidden, ICD offers a new ultrafast decay path typically on the femtosecond timescale where energy is exchanged with a neighboring atom leading to its ionization. Since its proposition~\cite{Cederbaum1997} and experimental confirmation~\cite{Marburger2003,Jahnke2004}, ICD has been observed in a wide variety of weakly-bound systems including, for example, He dimers~\cite{Sisourat2010,Havermeier2010} and water clusters~\cite{Jahnke2010,Mucke2010}. For reviews, see~\cite{Hergenhahn2011,Jahnke2015}.

Electron transfer mediated decay (ETMD), theoretically predicted~\cite{Zobeley2001} and recently experimentally observed~\cite{Sakai2011,Foerstel2011}, is another interatomic decay mechanism accessible in weakly-bound systems. Charge transfer (CT) from the neighbor to the ion releases energy which is utilized to either directly emit a second electron leading to double ionization, ETMD(2), or the released energy is transferred to a second neighbor leading to a single ionization of the two neighbors, ETMD(3). Importantly, ETMD is a much stronger decay channel than its radiative counterpart. Like in ICD, the original version of ETMD applies to ions which possess sufficient excess energy to ionize their neighbors. In sharp contrast to both cases, recently, a new variant of ETMD has been disclosed which does not require excess energy and can even be triggered from the ground state~\cite{Stumpf2013}. 

Recently, Stumpf et al.~\cite{Stumpf2014} predicted ETMD to dramatically enhance ($\sim$ 3 orders of magnitude) the single photon double ionization of a Mg atom in the vicinity of a He atom. In this case, ETMD proceeds by the initial ionization of a He atom followed by ETMD of the neighboring Mg atom yielding Mg$^{2+}$ and neutral He. Surprisingly, due to ETMD, the theoretical cross section for {\it double} ionization of Mg is even higher than that of direct {\it single} ionization and is comparable to that of He. Overall, the decay path and its predicted enhancement is not limited to Mg in He clusters, but can be applied to any embedded atoms or molecules which have a lower double ionization potential than the single ionization potential of the environment. Thus, the phenomenon is considered to be of quite general relevance and can be used in He droplets as a new pathway to the formation of doubly-ionized cold species which are difficult to form otherwise.

Here, we report on the first experimental observation of ETMD of particles embedded in superfluid He nanodroplets. Following the initial ionization of a He atom within the droplet, ETMD leads to double ionization of the embedded Mg clusters. The electron kinetic energy spectra reveal a low energy ETMD peak at about 1 eV agreeing well with theory. The ETMD mechanism turns out to be a dominant means to doubly ionize Mg clusters within the droplets allowing the investigation of the stability of doubly-ionized Mg clusters.

He droplets have widely been used as a cold, weakly perturbing matrices for studies in spectroscopy and chemical dynamics of embedded atoms and molecules~\cite{Toennies2004,Stienkemeier2006}. While typically the He environment is inert to the embedded species, when the droplet is excited or ionized the situation is completely different and the droplet becomes a highly reactive medium to the embedded species~\cite{Mudrich2014}. Even doubly-ionized dopants have recently been observed due to sequential collisions of metastable He atoms produced in a single droplet by electron bombardment~\cite{Schoebel2010}.

The experiment was performed using a mobile He droplet machine attached to an imaging photoelectron-photoion coincidence (PEPICO) detector at the GasPhase beamline of Elettra-Sincrotrone Trieste, Italy. The setup has been described in some detail earlier~\cite{OKeeffe2011,Buchta2013}, and only the significant points will be addressed here. In short, a beam of He nanodroplets is produced by continuously expanding pressurized (50 bar), high purity He out of a cryogenic nozzle with 5 $\mu$m diameter. Under these expansion conditions, the mean droplet sizes range from 10$^1$ to 10$^{11}$ He atoms per droplet~\cite{Toennies2004}. After passing a skimmer (0.4 mm) and a mechanical beam chopper used for discriminating the droplet beam signal from the He background, the droplets were doped using the “pick-up” technique~\cite{Gough1985} with an oven cell filled with Mg heated to generate partial pressures where 1-25 Mg atoms were attached to the droplets. While most atomic and molecular species become submerged into the interior of He nanodroplets, alkali earth atoms such as Mg remain weakly bound inside the surface layer~\cite{Barranco2006}. The He droplet beam next crosses the synchrotron beam at the focus of the PEPICO detector consisting of an ion time-of-flight detector and velocity map imaging detector operating in coincidence. With such a detection technique, one can record electron kinetic energy distributions in coincidence with a specific ion masses in multicoincidence mode~\cite{OKeeffe2011}. The kinetic energy distributions were reconstructed using a standard Abel inversion method~\cite{Garcia2004}. The photon energy was tuned by scanning the monochromator and gap of the undulator simultaneously with a typical step size of 20 meV and energy resolution $E/\Delta E\,\approx\,10^4$. The intensity of the radiation was monitored by a calibrated photodiode and all photon energy dependent ion and electron spectra shown in this work are normalized to this intensity signal.  

\begin{figure}
\begin{center}{
\includegraphics[width=0.5\textwidth]{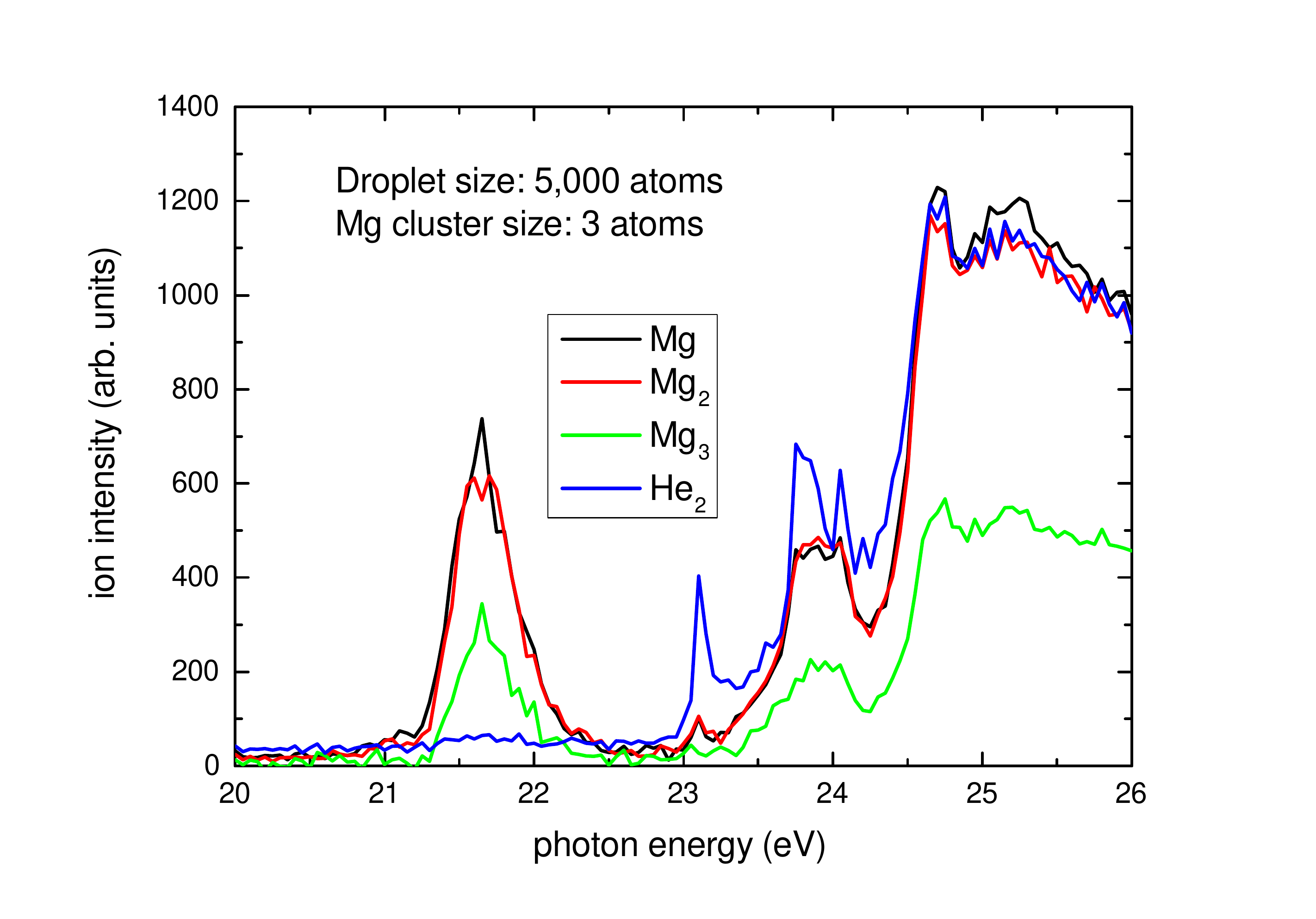}}
\caption{Ion signal intensity of Mg (black), Mg$_2$ (red), Mg$_3$ (green), and He$_2$ (blue) as a function of photon energy. The droplet size is 5\,000 He atoms with an average 
of 2-3 Mg atoms attached.}
\label{fig1}
\end{center}
\end{figure}

Fig.\,\ref{fig1} shows the ion signal intensities of Mg$^{+}$, Mg$_2^{+}$, Mg$_3^{+}$, and He$_2^{+}$ as a function of photon energy. The droplet size is 5\,000 He atoms with an average of 2-3 Mg atoms attached. The photon energy was tuned from 20 eV to 26 eV, which covers energies below the lowest dipole-allowed excitation energy (21.2 eV) to above the atomic ionization threshold (24.6 eV) of He droplets~\cite{Joppien1993}. The observed He ion signals are similar to previous synchrotron results~\cite{Buchta2013a,Froechtenicht1996} where for energies below about 23 eV (the adiabatic ionization potential of He droplets) the signal is nearly zero. At energies higher than 23 eV but below the ionization threshold of atomic He ($h\nu$ $\leq$ 24.6 eV), ionization occurs through He dimerization followed by autoionization~\cite{Froechtenicht1996}.

There are several mechanisms leading to single ionization of Mg atoms and clusters embedded in He droplets. First, direct single ionization above 7.6 eV and double ionization above 22.7 eV are possible~\cite{Kramida2014}. Additionally, the He environment opens new pathways to ionization. When the droplet is resonantly excited~\cite{Joppien1993}, ionization proceeds through Penning or ICD~\cite{Wang2008,Trinter2013} processes. When the droplet is ionized, CT can ionize the dopant. For high enough photon energies, electron impact ionization of the dopant is possible by photoelectrons produced in the initial ionization of the He droplet. All the processes above lead to the production of a single Mg$^{+}$ or Mg$_{n}^{+}$ ion by a single photon. A single photon can also doubly ionize Mg$_{n}$ via ETMD~\cite{Stumpf2014}. The thus produced  Mg$_{n}^{2+}$ can either fragment into two singly charged fragments or remain a stable dication. 

For the Mg ions below 21 eV photon energy, no signal was observed; therefore, direct ionization of Mg is negligible. Around 21.6 eV, there is a large peak in all three Mg ion signals in Fig.\,\ref{fig1} corresponding to ionization by Penning or ICD processes~\cite{Wang2008,Trinter2013} as the He atoms are excited to the droplet-equivalent of the 1s2p state~\cite{Joppien1993}. At higher photon energies, the Mg ion signals closely follow the He$_2^{+}$ ion signal pointing at He-mediated ionization of Mg. Previous experiments with dopants (alkali metals) which cannot undergo ETMD exhibit efficient dopant ionization at the excitation energies of the droplet. However, ionization of the dopant was comparatively weak at higher photon energies above the droplet's ionization threshold in contrast to the case of Mg presented here~\cite{Scheidemann1997,Buchta2013}. This is surprising considering that the cross section for resonant excitation of He~\cite{Buchta2013a} is three times higher than the ionization cross section near threshold~\cite{Marr1976} and that Mg is located close to the droplet's surface similar to alklali metals~\cite{Ren2007}. The question which arises is whether the strong enhancement is due to ETMD.

\begin{figure}
\begin{center}{
\includegraphics[width=0.5\textwidth]{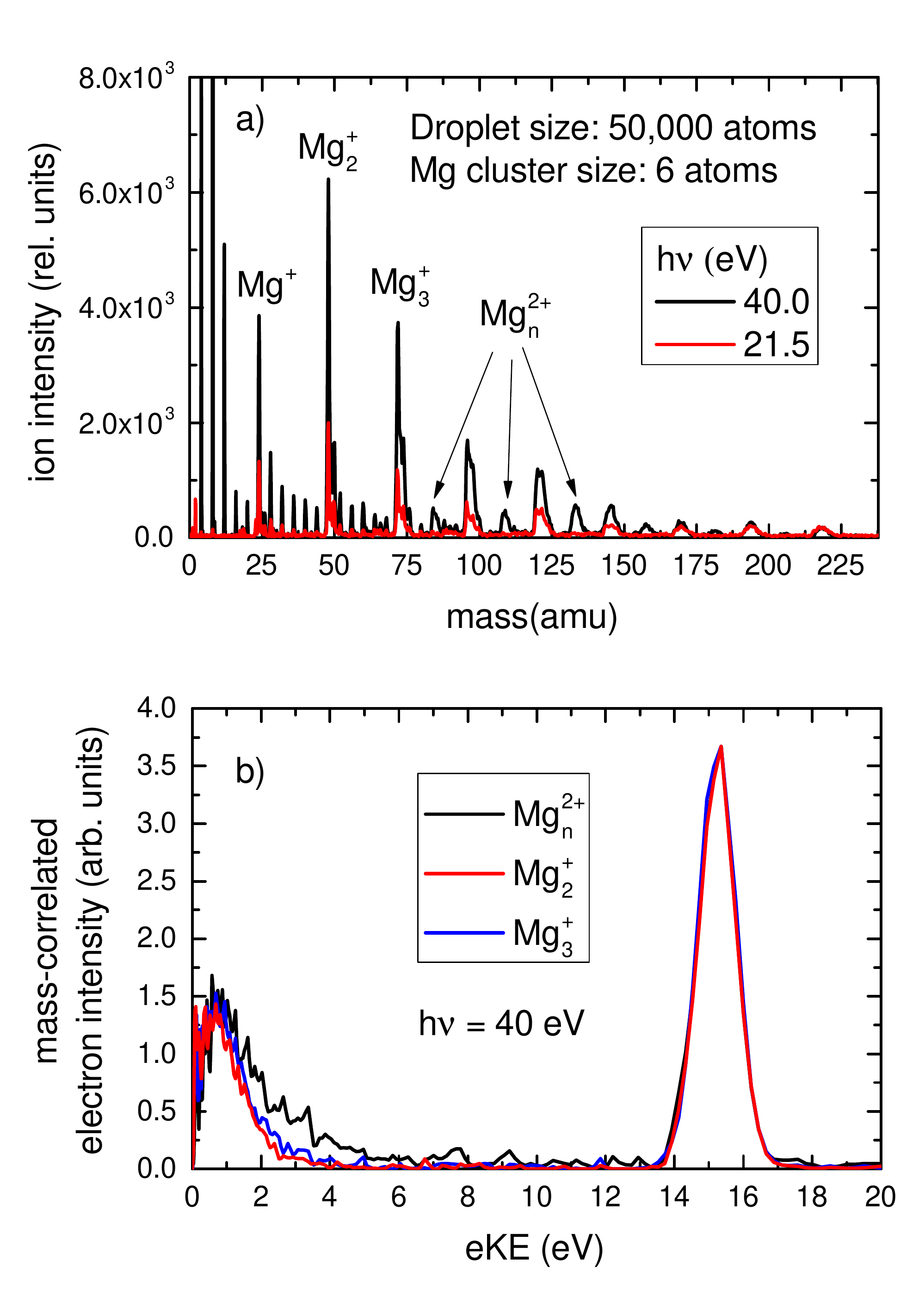}}
\caption{a) Mass spectra of He droplets doped with Mg clusters at photon energies of 40 eV (black) and 21.5 eV (red). The droplet size is 50\,000 He atoms with an average of 5-6 Mg atoms attached. b) Electron kinetic energy distributions measured in coincidence with single ions for Mg$_{n=7,9,11}^{2+}$ (black line), Mg$_2^+$ (red line), and Mg$_3^+$ (blue line).}
\label{fig2}
\end{center}
\end{figure}

To address the differences in the ionization mechanisms discussed above, we show in Fig.\,\ref{fig2} a) the mass spectra for photon energies of 40 eV (black line) and 21.5 eV (red line) for droplets consisting of 50\,000 He atoms with an average of 5-6 Mg atoms attached. For both energies, a large contribution of Mg ions is observed in the mass spectra, and, similar to Fig.\,\ref{fig1}, there are substantially higher signals above the ionization threshold(h$\nu$\,=\,40 eV). Here, Mg$^+$-He complexes are observed at multiples of 4 amu in the mass spectra following multiples of the Mg mass (24 amu).

Importantly, at higher masses, broad peaks at half-integer values of the mass/charge ratio appear. These are due to the formation of doubly-ionized Mg clusters with at least 5 atoms. The stability of doubly-ionized clusters has previously been studied~\cite{Reuse1990,Diederich2005} and it was experimentally shown that Mg clusters consisting of 5 atoms or more are sufficiently long-lived to be detected in a mass spectrometer. The signals in the mass spectrum corresponding to integer numbers of Mg atoms in Mg$_n^+$ may, of course also be due to Mg$_n^{2+}$. However, it is impossible to disentangle them from singly ionized clusters. The observation of doubly-ionized Mg clusters gives the first direct evidence of ETMD for this system. 

In order to identify the various ionization mechanisms, mass-correlated electron spectra are shown in Fig.\,\ref{fig2} b) for Mg$_n^{2+}$ (black line). As there was no difference between the electron spectra correlated to the various doubly-ionized clusters, they were combined to increase statistics. At 15.4 eV, one observes a large photoelectron peak resulting from the initial ionization of He ($h\nu$\,–\,E$_{i}$(He)\,=\,15.4 eV). The Mg$_n^{2+}$ peak at low energy is due to ETMD as the electrons emitted in this process lie in the observed energy range (see supplementary material and Ref.~\cite{Stumpf2014}) and there are no other mechanisms which produce an electron peak in this energy range. We shall argue below that ETMD is by far the dominant mechanism for producing doubly-ionized Mg clusters. Shown in Fig.\,\ref{fig2} b) are also the mass-correlated electron spectra for the Mg$_2^{+}$ (red line) and Mg$_3^{+}$ (blue line) ions. Surprisingly, they are similar to that of Mg$_n^{2+}$, exhibiting a photoelectron peak at 15.4 eV and a low-energy peak at about 1 eV. Therefore, not only is ETMD responsible for the doubly-ionized clusters in the mass spectra but could be a primary mechanism for the production of singly charged smaller clusters, which result upon fragmentation of doubly-ionized unstable clusters. There is a slight discrepancy between the spectra for Mg$_n^{2+}$ and  those of smaller clusters, Mg$_2^{+}$ and Mg$_3^{+}$; namely, the ETMD peak for Mg$_n^{2+}$ extends to higher energies. The additional energy for Mg$_n^{2+}$ clusters is due to the clusters not undergoing dissociation which requires additional energy. The supplementary material contains mass and electron spectra similar to those shown in Fig.\,\ref{fig2} for 5\,000 He atoms, but with an average of 2-3 Mg atoms attached. In this case, Mg$_5^{2+}$ can clearly be identified in the mass spectra.

\begin{figure}
\begin{center}{
\includegraphics[width=0.5\textwidth]{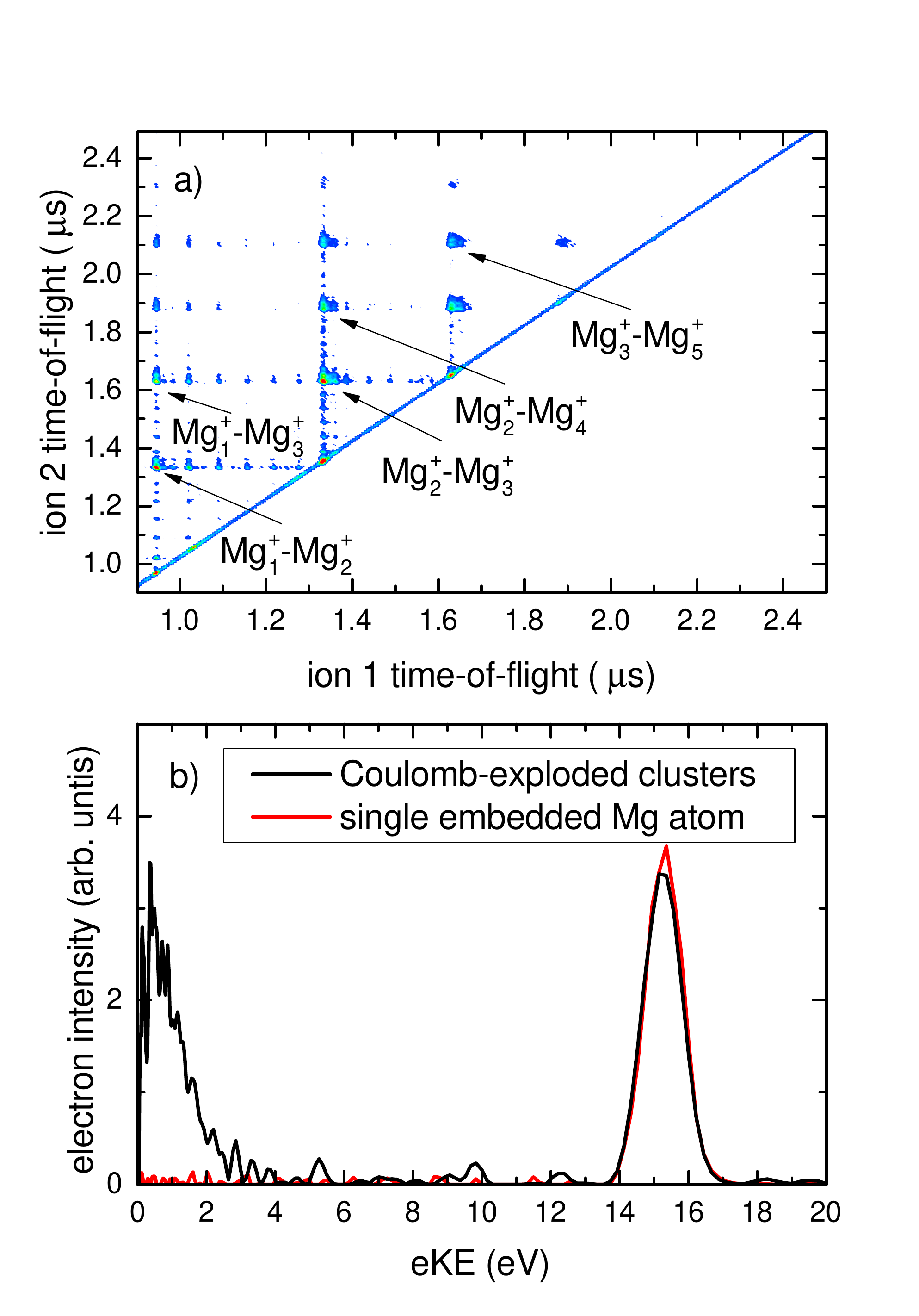}}
\caption{a) Ion-ion coincidence time-of-flight spectrum and b) electron spectrum correlated to Mg$_m^+$-Mg$_n^+$ ion-ion coincidences (black line). The photon energy is 40 eV. The droplet size is 50\,000 He atoms with an average of six Mg atoms attached. electron spectra correlated to a single Mg atom (red line) embedded in a droplet of size 5\,000 atoms.}
\label{fig3}
\end{center}
\end{figure}

Fig.\,\ref{fig3} a) shows the ion-ion coincidence time-of-flight spectrum for He droplets consisting of 50\,000 atoms doped with an average of 5-6 Mg atoms. The photon energy was 40 eV. As the flight times of the respective ions are symmetric, the coincidence map is folded along the axis of symmetry. The peak shapes observed in the spectra give information about the dissociation process~\cite{Eland1991}. Additionally, due to the dense He environment, the kinetic energy of Coulomb-exploded molecules embedded in He droplets is significantly damped. The coincidence map is centered around Coulomb-exploded Mg oligomers. Due to the dead time of the detector ($\approx$ 9 ns), there is a significant loss of statistics for ions of equal mass lying along the axis of symmetry. Therefore, the focus of our results is on Coulomb-exploded clusters of unequal masses,  Mg$_m^+$-$-$Mg$_n^+$ where $n\neq m$. Overall, the coincidence map reveals a rich spectrum of doubly-ionized clusters up to Mg$_9$ showing that many of the singly ionized Mg ions seen in Fig.\,\ref{fig2} a) stem from larger doubly-ionized clusters. 

Fig.\,\ref{fig3} b) shows the electron spectra correlated to ion-ion coincidences (black line). Since there was no substantial change (e.g. peak position and width or relative ratio of the ETMD to photoelectron peak) for the various Mg ion-ion pairs, the individual electron spectra were summed for all large heterogeneous ion-ion pairs shown in the coincidence map in Fig.\,\ref{fig3} a). Similar to the electron spectra shown in Fig.\,\ref{fig2} b), the ETMD peak is also centered at 0.9 eV. Here, the ratio of the integrated ETMD peak to the photoelectron peak is 81\% in this case, significantly higher than that shown in Fig.\,\ref{fig2} b) (35-50\% depending on the correlated ion) consistent that doubly-ionized clusters are a result of ETMD. The most likely reason that the ETMD to photoelectron peak ratio is less than unity is due to the large background signal of false coincidences from ionized He. Besides the photoelectron and ETMD peaks, there are no electron signals from other mechanisms, which highlights that ETMD is the dominant process for double ionization of Mg clusters. The ion-ion coincidence map and electron spectra for He droplets consisting of 5\,000 atoms doped with on average 2-3 Mg atoms are given in the supplementary material and give consistent results with those in Fig.\ref{fig3}. 

So far, we have solely focused on Mg clusters embedded in He nanodroplets where we have shown that the He environment dramatically enhances the double ionization of the cluster due to ETMD. Finally, we address the enhancement of the double ionization efficiency of a single Mg atom with a single He atom as a neighbor as investigated by Stumpf et al.~\cite{Stumpf2014}. We observed no doubly-ionized Mg atoms when only a single Mg atom is embedded in the nanodroplet; see supplementary material. The Mg$^+$-correlated electron spectra for a single Mg atom embedded in a droplet consisting of 5\,000 atoms is shown in Fig.\,\ref{fig3} b). In this case, the ETMD electron peak was absent suggesting that ionization proceeds exclusively through CT. Thus, ETMD appears to be inactive for single Mg atoms attached to He droplets.

We attribute this to the ultrafast formation of He$_2^{+}$ in the droplet, which is predicted to occur in 60-80 fs after the initial ionization~\cite{Buchta2013}. Our calculations show clearly that the ETMD channel for a single Mg atom is closed once He$_2^{+}$ is formed. Additionally included in the supplementary material are theoretical calculations of the ETMD electron kinetic energy for the most likely configurations between He$_2^{+}$ and Mg trimer where ETMD can still occur. It is yet inconclusive whether ETMD can occur for Mg dimers. 

In conclusion, electron transfer mediated decay was observed for Mg clusters embedded in He droplets. This decay channel was shown to be a dominant ionization mechanism for energies above the ionization threshold of He. For clusters of 5 Mg atoms and greater, stable, doubly-ionized Mg clusters were observed after ETMD. For single Mg atoms embedded in nanodroplets, the ETMD channel is closed due to the ultrafast formation of an equilibrated He dimer ion. In general, ETMD offers a novel method for producing doubly-ionized systems if the environment has a higher single ionization threshold.


\begin{thebibliography}{38}%
\makeatletter
\providecommand \@ifxundefined [1]{%
 \@ifx{#1\undefined}
}%
\providecommand \@ifnum [1]{%
 \ifnum #1\expandafter \@firstoftwo
 \else \expandafter \@secondoftwo
 \fi
}%
\providecommand \@ifx [1]{%
 \ifx #1\expandafter \@firstoftwo
 \else \expandafter \@secondoftwo
 \fi
}%
\providecommand \natexlab [1]{#1}%
\providecommand \enquote  [1]{``#1''}%
\providecommand \bibnamefont  [1]{#1}%
\providecommand \bibfnamefont [1]{#1}%
\providecommand \citenamefont [1]{#1}%
\providecommand \href@noop [0]{\@secondoftwo}%
\providecommand \href [0]{\begingroup \@sanitize@url \@href}%
\providecommand \@href[1]{\@@startlink{#1}\@@href}%
\providecommand \@@href[1]{\endgroup#1\@@endlink}%
\providecommand \@sanitize@url [0]{\catcode `\\12\catcode `\$12\catcode
  `\&12\catcode `\#12\catcode `\^12\catcode `\_12\catcode `\%12\relax}%
\providecommand \@@startlink[1]{}%
\providecommand \@@endlink[0]{}%
\providecommand \url  [0]{\begingroup\@sanitize@url \@url }%
\providecommand \@url [1]{\endgroup\@href {#1}{\urlprefix }}%
\providecommand \urlprefix  [0]{URL }%
\providecommand \Eprint [0]{\href }%
\providecommand \doibase [0]{http://dx.doi.org/}%
\providecommand \selectlanguage [0]{\@gobble}%
\providecommand \bibinfo  [0]{\@secondoftwo}%
\providecommand \bibfield  [0]{\@secondoftwo}%
\providecommand \translation [1]{[#1]}%
\providecommand \BibitemOpen [0]{}%
\providecommand \bibitemStop [0]{}%
\providecommand \bibitemNoStop [0]{.\EOS\space}%
\providecommand \EOS [0]{\spacefactor3000\relax}%
\providecommand \BibitemShut  [1]{\csname bibitem#1\endcsname}%
\let\auto@bib@innerbib\@empty
\bibitem [{\citenamefont {Pattard}\ \emph {et~al.}(2003)\citenamefont
  {Pattard}, \citenamefont {Schneider},\ and\ \citenamefont
  {Rost}}]{Pattard2003}%
  \BibitemOpen
  \bibfield  {author} {\bibinfo {author} {\bibfnamefont {T.}~\bibnamefont
  {Pattard}}, \bibinfo {author} {\bibfnamefont {T.}~\bibnamefont {Schneider}},
  \ and\ \bibinfo {author} {\bibfnamefont {J.}~\bibnamefont {Rost}},\
  }\href@noop {} {\bibfield  {journal} {\bibinfo  {journal} {J. Phys. B: At.,
  Mol. Opt. Phys.}\ }\textbf {\bibinfo {volume} {36}},\ \bibinfo {pages} {L189}
  (\bibinfo {year} {2003})}\BibitemShut {NoStop}%
\bibitem [{\citenamefont {Russek}\ and\ \citenamefont
  {Mehlhorn}(1986)}]{Russek1986}%
  \BibitemOpen
  \bibfield  {author} {\bibinfo {author} {\bibfnamefont {A.}~\bibnamefont
  {Russek}}\ and\ \bibinfo {author} {\bibfnamefont {W.}~\bibnamefont
  {Mehlhorn}},\ }\href@noop {} {\bibfield  {journal} {\bibinfo  {journal} {J.
  Phys. B: At., Mol. Opt. Phys.}\ }\textbf {\bibinfo {volume} {19}},\ \bibinfo
  {pages} {911} (\bibinfo {year} {1986})}\BibitemShut {NoStop}%
\bibitem [{\citenamefont {Madden}\ and\ \citenamefont
  {Codling}(1963)}]{Madden1963}%
  \BibitemOpen
  \bibfield  {author} {\bibinfo {author} {\bibfnamefont {R.}~\bibnamefont
  {Madden}}\ and\ \bibinfo {author} {\bibfnamefont {K.}~\bibnamefont
  {Codling}},\ }\href@noop {} {\bibfield  {journal} {\bibinfo  {journal} {Phys.
  Rev. Lett.}\ }\textbf {\bibinfo {volume} {10}},\ \bibinfo {pages} {516}
  (\bibinfo {year} {1963})}\BibitemShut {NoStop}%
\bibitem [{\citenamefont {Cederbaum}\ \emph {et~al.}(1997)\citenamefont
  {Cederbaum}, \citenamefont {Zobeley},\ and\ \citenamefont
  {Tarantelli}}]{Cederbaum1997}%
  \BibitemOpen
  \bibfield  {author} {\bibinfo {author} {\bibfnamefont {L.~S.}\ \bibnamefont
  {Cederbaum}}, \bibinfo {author} {\bibfnamefont {J.}~\bibnamefont {Zobeley}},
  \ and\ \bibinfo {author} {\bibfnamefont {F.}~\bibnamefont {Tarantelli}},\
  }\href {\doibase 10.1103/PhysRevLett.79.4778} {\bibfield  {journal} {\bibinfo
   {journal} {Phys. Rev. Lett.}\ }\textbf {\bibinfo {volume} {79}},\ \bibinfo
  {pages} {4778} (\bibinfo {year} {1997})}\BibitemShut {NoStop}%
\bibitem [{\citenamefont {Marburger}\ \emph {et~al.}(2003)\citenamefont
  {Marburger}, \citenamefont {Kugeler}, \citenamefont {Hergenhahn},\ and\
  \citenamefont {M\"oller}}]{Marburger2003}%
  \BibitemOpen
  \bibfield  {author} {\bibinfo {author} {\bibfnamefont {S.}~\bibnamefont
  {Marburger}}, \bibinfo {author} {\bibfnamefont {O.}~\bibnamefont {Kugeler}},
  \bibinfo {author} {\bibfnamefont {U.}~\bibnamefont {Hergenhahn}}, \ and\
  \bibinfo {author} {\bibfnamefont {T.}~\bibnamefont {M\"oller}},\ }\href
  {\doibase 10.1103/PhysRevLett.90.203401} {\bibfield  {journal} {\bibinfo
  {journal} {Phys. Rev. Lett.}\ }\textbf {\bibinfo {volume} {90}},\ \bibinfo
  {pages} {203401} (\bibinfo {year} {2003})}\BibitemShut {NoStop}%
\bibitem [{\citenamefont {Jahnke}\ \emph {et~al.}(2004)\citenamefont {Jahnke}
  \emph {et~al.}}]{Jahnke2004}%
  \BibitemOpen
  \bibfield  {author} {\bibinfo {author} {\bibfnamefont {T.}~\bibnamefont
  {Jahnke}} \emph {et~al.},\ }\href {\doibase 10.1103/PhysRevLett.93.163401}
  {\bibfield  {journal} {\bibinfo  {journal} {Phys. Rev. Lett.}\ }\textbf
  {\bibinfo {volume} {93}},\ \bibinfo {pages} {163401} (\bibinfo {year}
  {2004})}\BibitemShut {NoStop}%
\bibitem [{\citenamefont {Sisourat}\ \emph {et~al.}(2010)\citenamefont
  {Sisourat} \emph {et~al.}}]{Sisourat2010}%
  \BibitemOpen
  \bibfield  {author} {\bibinfo {author} {\bibfnamefont {N.}~\bibnamefont
  {Sisourat}} \emph {et~al.},\ }\href@noop {} {\bibfield  {journal} {\bibinfo
  {journal} {Nat. Phys.}\ }\textbf {\bibinfo {volume} {6}},\ \bibinfo {pages}
  {508} (\bibinfo {year} {2010})}\BibitemShut {NoStop}%
\bibitem [{\citenamefont {Havermeier}\ \emph {et~al.}(2010)\citenamefont
  {Havermeier} \emph {et~al.}}]{Havermeier2010}%
  \BibitemOpen
  \bibfield  {author} {\bibinfo {author} {\bibfnamefont {T.}~\bibnamefont
  {Havermeier}} \emph {et~al.},\ }\href@noop {} {\bibfield  {journal} {\bibinfo
   {journal} {Phys. Rev. Lett.}\ }\textbf {\bibinfo {volume} {104}},\ \bibinfo
  {pages} {133401} (\bibinfo {year} {2010})}\BibitemShut {NoStop}%
\bibitem [{\citenamefont {Jahnke}\ \emph {et~al.}(2010)\citenamefont {Jahnke}
  \emph {et~al.}}]{Jahnke2010}%
  \BibitemOpen
  \bibfield  {author} {\bibinfo {author} {\bibfnamefont {T.}~\bibnamefont
  {Jahnke}} \emph {et~al.},\ }\href@noop {} {\bibfield  {journal} {\bibinfo
  {journal} {Nat. Phys.}\ }\textbf {\bibinfo {volume} {6}},\ \bibinfo {pages}
  {139} (\bibinfo {year} {2010})}\BibitemShut {NoStop}%
\bibitem [{\citenamefont {Mucke}\ \emph {et~al.}(2010)\citenamefont {Mucke}
  \emph {et~al.}}]{Mucke2010}%
  \BibitemOpen
  \bibfield  {author} {\bibinfo {author} {\bibfnamefont {M.}~\bibnamefont
  {Mucke}} \emph {et~al.},\ }\href@noop {} {\bibfield  {journal} {\bibinfo
  {journal} {Nat. Phys.}\ }\textbf {\bibinfo {volume} {6}},\ \bibinfo {pages}
  {143} (\bibinfo {year} {2010})}\BibitemShut {NoStop}%
\bibitem [{\citenamefont {Hergenhahn}(2011)}]{Hergenhahn2011}%
  \BibitemOpen
  \bibfield  {author} {\bibinfo {author} {\bibfnamefont {U.}~\bibnamefont
  {Hergenhahn}},\ }\href@noop {} {\bibfield  {journal} {\bibinfo  {journal} {J.
  Electron. Spectrosc. Relat. Phenom.}\ }\textbf {\bibinfo {volume} {184}},\
  \bibinfo {pages} {78} (\bibinfo {year} {2011})}\BibitemShut {NoStop}%
\bibitem [{\citenamefont {Jahnke}(2015)}]{Jahnke2015}%
  \BibitemOpen
  \bibfield  {author} {\bibinfo {author} {\bibfnamefont {T.}~\bibnamefont
  {Jahnke}},\ }\href@noop {} {\bibfield  {journal} {\bibinfo  {journal} {J.
  Phys. B: At., Mol. Opt. Phys.}\ }\textbf {\bibinfo {volume} {48}},\ \bibinfo
  {pages} {082001} (\bibinfo {year} {2015})}\BibitemShut {NoStop}%
\bibitem [{\citenamefont {Zobeley}\ \emph {et~al.}(2001)\citenamefont
  {Zobeley}, \citenamefont {Santra},\ and\ \citenamefont
  {Cederbaum}}]{Zobeley2001}%
  \BibitemOpen
  \bibfield  {author} {\bibinfo {author} {\bibfnamefont {J.}~\bibnamefont
  {Zobeley}}, \bibinfo {author} {\bibfnamefont {R.}~\bibnamefont {Santra}}, \
  and\ \bibinfo {author} {\bibfnamefont {L.~S.}\ \bibnamefont {Cederbaum}},\
  }\href@noop {} {\bibfield  {journal} {\bibinfo  {journal} {J. Chem. Phys.}\
  }\textbf {\bibinfo {volume} {115}},\ \bibinfo {pages} {5076} (\bibinfo {year}
  {2001})}\BibitemShut {NoStop}%
\bibitem [{\citenamefont {Sakai}\ \emph {et~al.}(2011)\citenamefont {Sakai}
  \emph {et~al.}}]{Sakai2011}%
  \BibitemOpen
  \bibfield  {author} {\bibinfo {author} {\bibfnamefont {K.}~\bibnamefont
  {Sakai}} \emph {et~al.},\ }\href@noop {} {\bibfield  {journal} {\bibinfo
  {journal} {Phys. Rev. Lett.}\ }\textbf {\bibinfo {volume} {106}},\ \bibinfo
  {pages} {033401} (\bibinfo {year} {2011})}\BibitemShut {NoStop}%
\bibitem [{\citenamefont {F{\"o}rstel}\ \emph {et~al.}(2011)\citenamefont
  {F{\"o}rstel}, \citenamefont {Mucke}, \citenamefont {Arion}, \citenamefont
  {Bradshaw},\ and\ \citenamefont {Hergenhahn}}]{Foerstel2011}%
  \BibitemOpen
  \bibfield  {author} {\bibinfo {author} {\bibfnamefont {M.}~\bibnamefont
  {F{\"o}rstel}}, \bibinfo {author} {\bibfnamefont {M.}~\bibnamefont {Mucke}},
  \bibinfo {author} {\bibfnamefont {T.}~\bibnamefont {Arion}}, \bibinfo
  {author} {\bibfnamefont {A.~M.}\ \bibnamefont {Bradshaw}}, \ and\ \bibinfo
  {author} {\bibfnamefont {U.}~\bibnamefont {Hergenhahn}},\ }\href@noop {}
  {\bibfield  {journal} {\bibinfo  {journal} {Phys. Rev. Lett.}\ }\textbf
  {\bibinfo {volume} {106}},\ \bibinfo {pages} {033402} (\bibinfo {year}
  {2011})}\BibitemShut {NoStop}%
\bibitem [{\citenamefont {Stumpf}\ \emph {et~al.}(2013)\citenamefont {Stumpf},
  \citenamefont {Koloren\ifmmode~\check{c}\else \v{c}\fi{}}, \citenamefont
  {Gokhberg},\ and\ \citenamefont {Cederbaum}}]{Stumpf2013}%
  \BibitemOpen
  \bibfield  {author} {\bibinfo {author} {\bibfnamefont {V.}~\bibnamefont
  {Stumpf}}, \bibinfo {author} {\bibfnamefont {P.}~\bibnamefont
  {Koloren\ifmmode~\check{c}\else \v{c}\fi{}}}, \bibinfo {author}
  {\bibfnamefont {K.}~\bibnamefont {Gokhberg}}, \ and\ \bibinfo {author}
  {\bibfnamefont {L.~S.}\ \bibnamefont {Cederbaum}},\ }\href {\doibase
  10.1103/PhysRevLett.110.258302} {\bibfield  {journal} {\bibinfo  {journal}
  {Phys. Rev. Lett.}\ }\textbf {\bibinfo {volume} {110}},\ \bibinfo {pages}
  {258302} (\bibinfo {year} {2013})}\BibitemShut {NoStop}%
\bibitem [{\citenamefont {Stumpf}\ \emph {et~al.}(2014)\citenamefont {Stumpf},
  \citenamefont {Kryzhevoi}, \citenamefont {Gokhberg},\ and\ \citenamefont
  {Cederbaum}}]{Stumpf2014}%
  \BibitemOpen
  \bibfield  {author} {\bibinfo {author} {\bibfnamefont {V.}~\bibnamefont
  {Stumpf}}, \bibinfo {author} {\bibfnamefont {N.}~\bibnamefont {Kryzhevoi}},
  \bibinfo {author} {\bibfnamefont {K.}~\bibnamefont {Gokhberg}}, \ and\
  \bibinfo {author} {\bibfnamefont {L.}~\bibnamefont {Cederbaum}},\ }\href@noop
  {} {\bibfield  {journal} {\bibinfo  {journal} {Phys. Rev. Lett.}\ }\textbf
  {\bibinfo {volume} {112}},\ \bibinfo {pages} {193001} (\bibinfo {year}
  {2014})}\BibitemShut {NoStop}%
\bibitem [{\citenamefont {Toennies}\ and\ \citenamefont
  {Vilesov}(2004)}]{Toennies2004}%
  \BibitemOpen
  \bibfield  {author} {\bibinfo {author} {\bibfnamefont {J.~P.}\ \bibnamefont
  {Toennies}}\ and\ \bibinfo {author} {\bibfnamefont {A.~F.}\ \bibnamefont
  {Vilesov}},\ }\href@noop {} {\bibfield  {journal} {\bibinfo  {journal}
  {Angew. Chem. Int. Ed.}\ }\textbf {\bibinfo {volume} {43}},\ \bibinfo {pages}
  {2622} (\bibinfo {year} {2004})}\BibitemShut {NoStop}%
\bibitem [{\citenamefont {Stienkemeier}\ and\ \citenamefont
  {Lehmann}(2006)}]{Stienkemeier2006}%
  \BibitemOpen
  \bibfield  {author} {\bibinfo {author} {\bibfnamefont {F.}~\bibnamefont
  {Stienkemeier}}\ and\ \bibinfo {author} {\bibfnamefont {K.~K.}\ \bibnamefont
  {Lehmann}},\ }\href@noop {} {\bibfield  {journal} {\bibinfo  {journal} {J.
  Phys. B: At., Mol. Opt. Phys.}\ }\textbf {\bibinfo {volume} {39}},\ \bibinfo
  {pages} {R127} (\bibinfo {year} {2006})}\BibitemShut {NoStop}%
\bibitem [{\citenamefont {Mudrich}\ and\ \citenamefont
  {Stienkemeier}(2014)}]{Mudrich2014}%
  \BibitemOpen
  \bibfield  {author} {\bibinfo {author} {\bibfnamefont {M.}~\bibnamefont
  {Mudrich}}\ and\ \bibinfo {author} {\bibfnamefont {F.}~\bibnamefont
  {Stienkemeier}},\ }\href@noop {} {\bibfield  {journal} {\bibinfo  {journal}
  {Int. Rev. Phys. Chem.}\ }\textbf {\bibinfo {volume} {33}},\ \bibinfo {pages}
  {301} (\bibinfo {year} {2014})}\BibitemShut {NoStop}%
\bibitem [{\citenamefont {Sch\"obel}\ \emph {et~al.}(2010)\citenamefont
  {Sch\"obel} \emph {et~al.}}]{Schoebel2010}%
  \BibitemOpen
  \bibfield  {author} {\bibinfo {author} {\bibfnamefont {H.}~\bibnamefont
  {Sch\"obel}} \emph {et~al.},\ }\href {\doibase
  10.1103/PhysRevLett.105.243402} {\bibfield  {journal} {\bibinfo  {journal}
  {Phys. Rev. Lett.}\ }\textbf {\bibinfo {volume} {105}},\ \bibinfo {pages}
  {243402} (\bibinfo {year} {2010})}\BibitemShut {NoStop}%
\bibitem [{\citenamefont {O'Keeffe}\ \emph {et~al.}(2011)\citenamefont
  {O'Keeffe} \emph {et~al.}}]{OKeeffe2011}%
  \BibitemOpen
  \bibfield  {author} {\bibinfo {author} {\bibfnamefont {P.}~\bibnamefont
  {O'Keeffe}} \emph {et~al.},\ }\href@noop {} {\bibfield  {journal} {\bibinfo
  {journal} {Rev. of Sci. Instrum.}\ }\textbf {\bibinfo {volume} {82}},\
  \bibinfo {pages} {033109} (\bibinfo {year} {2011})}\BibitemShut {NoStop}%
\bibitem [{\citenamefont {Buchta}\ \emph
  {et~al.}(2013{\natexlab{a}})\citenamefont {Buchta} \emph
  {et~al.}}]{Buchta2013}%
  \BibitemOpen
  \bibfield  {author} {\bibinfo {author} {\bibfnamefont {D.}~\bibnamefont
  {Buchta}} \emph {et~al.},\ }\href@noop {} {\bibfield  {journal} {\bibinfo
  {journal} {J. Phys. Chem. A}\ }\textbf {\bibinfo {volume} {117}},\ \bibinfo
  {pages} {4394} (\bibinfo {year} {2013}{\natexlab{a}})}\BibitemShut {NoStop}%
\bibitem [{\citenamefont {Gough}\ \emph {et~al.}(1985)\citenamefont {Gough},
  \citenamefont {Mengel}, \citenamefont {Rowntree},\ and\ \citenamefont
  {Scoles}}]{Gough1985}%
  \BibitemOpen
  \bibfield  {author} {\bibinfo {author} {\bibfnamefont {T.}~\bibnamefont
  {Gough}}, \bibinfo {author} {\bibfnamefont {M.}~\bibnamefont {Mengel}},
  \bibinfo {author} {\bibfnamefont {P.}~\bibnamefont {Rowntree}}, \ and\
  \bibinfo {author} {\bibfnamefont {G.}~\bibnamefont {Scoles}},\ }\href@noop {}
  {\bibfield  {journal} {\bibinfo  {journal} {J. Chem. Phys.}\ }\textbf
  {\bibinfo {volume} {83}},\ \bibinfo {pages} {4958} (\bibinfo {year}
  {1985})}\BibitemShut {NoStop}%
\bibitem [{\citenamefont {Barranco}\ \emph {et~al.}(2006)\citenamefont
  {Barranco}, \citenamefont {Guardiola}, \citenamefont {Hern{\'a}ndez},
  \citenamefont {Mayol}, \citenamefont {Navarro},\ and\ \citenamefont
  {Pi}}]{Barranco2006}%
  \BibitemOpen
  \bibfield  {author} {\bibinfo {author} {\bibfnamefont {M.}~\bibnamefont
  {Barranco}}, \bibinfo {author} {\bibfnamefont {R.}~\bibnamefont {Guardiola}},
  \bibinfo {author} {\bibfnamefont {S.}~\bibnamefont {Hern{\'a}ndez}}, \bibinfo
  {author} {\bibfnamefont {R.}~\bibnamefont {Mayol}}, \bibinfo {author}
  {\bibfnamefont {J.}~\bibnamefont {Navarro}}, \ and\ \bibinfo {author}
  {\bibfnamefont {M.}~\bibnamefont {Pi}},\ }\href@noop {} {\bibfield  {journal}
  {\bibinfo  {journal} {J. Low Temp. Phys.}\ }\textbf {\bibinfo {volume}
  {142}},\ \bibinfo {pages} {1} (\bibinfo {year} {2006})}\BibitemShut {NoStop}%
\bibitem [{\citenamefont {Garcia}\ \emph {et~al.}(2004)\citenamefont {Garcia},
  \citenamefont {Nahon},\ and\ \citenamefont {Powis}}]{Garcia2004}%
  \BibitemOpen
  \bibfield  {author} {\bibinfo {author} {\bibfnamefont {G.~A.}\ \bibnamefont
  {Garcia}}, \bibinfo {author} {\bibfnamefont {L.}~\bibnamefont {Nahon}}, \
  and\ \bibinfo {author} {\bibfnamefont {I.}~\bibnamefont {Powis}},\
  }\href@noop {} {\bibfield  {journal} {\bibinfo  {journal} {Rev. of Sci.
  Instrum.}\ }\textbf {\bibinfo {volume} {75}},\ \bibinfo {pages} {4989}
  (\bibinfo {year} {2004})}\BibitemShut {NoStop}%
\bibitem [{\citenamefont {Joppien}\ \emph {et~al.}(1993)\citenamefont
  {Joppien}, \citenamefont {Karnbach},\ and\ \citenamefont
  {M{\"o}ller}}]{Joppien1993}%
  \BibitemOpen
  \bibfield  {author} {\bibinfo {author} {\bibfnamefont {M.}~\bibnamefont
  {Joppien}}, \bibinfo {author} {\bibfnamefont {R.}~\bibnamefont {Karnbach}}, \
  and\ \bibinfo {author} {\bibfnamefont {T.}~\bibnamefont {M{\"o}ller}},\
  }\href@noop {} {\bibfield  {journal} {\bibinfo  {journal} {Phys. Rev. Lett.}\
  }\textbf {\bibinfo {volume} {71}},\ \bibinfo {pages} {2654} (\bibinfo {year}
  {1993})}\BibitemShut {NoStop}%
\bibitem [{\citenamefont {Buchta}\ \emph
  {et~al.}(2013{\natexlab{b}})\citenamefont {Buchta} \emph
  {et~al.}}]{Buchta2013a}%
  \BibitemOpen
  \bibfield  {author} {\bibinfo {author} {\bibfnamefont {D.}~\bibnamefont
  {Buchta}} \emph {et~al.},\ }\href@noop {} {\bibfield  {journal} {\bibinfo
  {journal} {J. Chem. Phys.}\ }\textbf {\bibinfo {volume} {139}},\ \bibinfo
  {pages} {084301} (\bibinfo {year} {2013}{\natexlab{b}})}\BibitemShut
  {NoStop}%
\bibitem [{\citenamefont {Fr{\"o}chtenicht}\ \emph {et~al.}(1996)\citenamefont
  {Fr{\"o}chtenicht}, \citenamefont {Henne}, \citenamefont {Toennies},
  \citenamefont {Ding}, \citenamefont {Fieber-Erdmann},\ and\ \citenamefont
  {Drewello}}]{Froechtenicht1996}%
  \BibitemOpen
  \bibfield  {author} {\bibinfo {author} {\bibfnamefont {R.}~\bibnamefont
  {Fr{\"o}chtenicht}}, \bibinfo {author} {\bibfnamefont {U.}~\bibnamefont
  {Henne}}, \bibinfo {author} {\bibfnamefont {J.~P.}\ \bibnamefont {Toennies}},
  \bibinfo {author} {\bibfnamefont {A.}~\bibnamefont {Ding}}, \bibinfo {author}
  {\bibfnamefont {M.}~\bibnamefont {Fieber-Erdmann}}, \ and\ \bibinfo {author}
  {\bibfnamefont {T.}~\bibnamefont {Drewello}},\ }\href@noop {} {\bibfield
  {journal} {\bibinfo  {journal} {J. Chem. Phys.}\ }\textbf {\bibinfo {volume}
  {104}},\ \bibinfo {pages} {2548} (\bibinfo {year} {1996})}\BibitemShut
  {NoStop}%
\bibitem [{\citenamefont {Kramida}\ \emph {et~al.}(2014)\citenamefont
  {Kramida}, \citenamefont {{Yu.~Ralchenko}}, \citenamefont {Reader},\ and\
  \citenamefont {{and NIST ASD Team}}}]{Kramida2014}%
  \BibitemOpen
  \bibfield  {author} {\bibinfo {author} {\bibfnamefont {A.}~\bibnamefont
  {Kramida}}, \bibinfo {author} {\bibnamefont {{Yu.~Ralchenko}}}, \bibinfo
  {author} {\bibfnamefont {J.}~\bibnamefont {Reader}}, \ and\ \bibinfo {author}
  {\bibnamefont {{and NIST ASD Team}}},\ }\href@noop {} {}\bibinfo
  {howpublished} {{NIST Atomic Spectra Database (ver. 5.2), [Online].
  Available: {\tt{http://physics.nist.gov/asd}} [2015, July 14]. National
  Institute of Standards and Technology, Gaithersburg, MD.}} (\bibinfo {year}
  {2014})\BibitemShut {NoStop}%
\bibitem [{\citenamefont {Wang}\ \emph {et~al.}(2008)\citenamefont {Wang} \emph
  {et~al.}}]{Wang2008}%
  \BibitemOpen
  \bibfield  {author} {\bibinfo {author} {\bibfnamefont {C.~C.}\ \bibnamefont
  {Wang}} \emph {et~al.},\ }\href@noop {} {\bibfield  {journal} {\bibinfo
  {journal} {J. Phys. Chem. A}\ }\textbf {\bibinfo {volume} {112}},\ \bibinfo
  {pages} {9356} (\bibinfo {year} {2008})}\BibitemShut {NoStop}%
\bibitem [{\citenamefont {Trinter}\ \emph {et~al.}(2013)\citenamefont {Trinter}
  \emph {et~al.}}]{Trinter2013}%
  \BibitemOpen
  \bibfield  {author} {\bibinfo {author} {\bibfnamefont {F.}~\bibnamefont
  {Trinter}} \emph {et~al.},\ }\href {\doibase 10.1103/PhysRevLett.111.233004}
  {\bibfield  {journal} {\bibinfo  {journal} {Phys. Rev. Lett.}\ }\textbf
  {\bibinfo {volume} {111}},\ \bibinfo {pages} {233004} (\bibinfo {year}
  {2013})}\BibitemShut {NoStop}%
\bibitem [{\citenamefont {Scheidemann}\ \emph {et~al.}(1997)\citenamefont
  {Scheidemann}, \citenamefont {Kresin},\ and\ \citenamefont
  {Hess}}]{Scheidemann1997}%
  \BibitemOpen
  \bibfield  {author} {\bibinfo {author} {\bibfnamefont {A.~A.}\ \bibnamefont
  {Scheidemann}}, \bibinfo {author} {\bibfnamefont {V.~V.}\ \bibnamefont
  {Kresin}}, \ and\ \bibinfo {author} {\bibfnamefont {H.}~\bibnamefont
  {Hess}},\ }\href@noop {} {\bibfield  {journal} {\bibinfo  {journal} {J. Chem.
  Phys.}\ }\textbf {\bibinfo {volume} {107}},\ \bibinfo {pages} {2839}
  (\bibinfo {year} {1997})}\BibitemShut {NoStop}%
\bibitem [{\citenamefont {Marr}\ and\ \citenamefont {West}(1976)}]{Marr1976}%
  \BibitemOpen
  \bibfield  {author} {\bibinfo {author} {\bibfnamefont {G.}~\bibnamefont
  {Marr}}\ and\ \bibinfo {author} {\bibfnamefont {J.}~\bibnamefont {West}},\
  }\href@noop {} {\bibfield  {journal} {\bibinfo  {journal} {At. Data Nucl.
  Data Tables}\ }\textbf {\bibinfo {volume} {18}},\ \bibinfo {pages} {497}
  (\bibinfo {year} {1976})}\BibitemShut {NoStop}%
\bibitem [{\citenamefont {Ren}\ and\ \citenamefont {Kresin}(2007)}]{Ren2007}%
  \BibitemOpen
  \bibfield  {author} {\bibinfo {author} {\bibfnamefont {Y.}~\bibnamefont
  {Ren}}\ and\ \bibinfo {author} {\bibfnamefont {V.~V.}\ \bibnamefont
  {Kresin}},\ }\href {\doibase 10.1103/PhysRevA.76.043204} {\bibfield
  {journal} {\bibinfo  {journal} {Phys. Rev. A}\ }\textbf {\bibinfo {volume}
  {76}},\ \bibinfo {pages} {043204} (\bibinfo {year} {2007})}\BibitemShut
  {NoStop}%
\bibitem [{\citenamefont {Reuse}\ \emph {et~al.}(1990)\citenamefont {Reuse},
  \citenamefont {Khanna}, \citenamefont {de~Coulon},\ and\ \citenamefont
  {Buttet}}]{Reuse1990}%
  \BibitemOpen
  \bibfield  {author} {\bibinfo {author} {\bibfnamefont {F.}~\bibnamefont
  {Reuse}}, \bibinfo {author} {\bibfnamefont {S.~N.}\ \bibnamefont {Khanna}},
  \bibinfo {author} {\bibfnamefont {V.}~\bibnamefont {de~Coulon}}, \ and\
  \bibinfo {author} {\bibfnamefont {J.}~\bibnamefont {Buttet}},\ }\href
  {\doibase 10.1103/PhysRevB.41.11743} {\bibfield  {journal} {\bibinfo
  {journal} {Phys. Rev. B}\ }\textbf {\bibinfo {volume} {41}},\ \bibinfo
  {pages} {11743} (\bibinfo {year} {1990})}\BibitemShut {NoStop}%
\bibitem [{\citenamefont {Diederich}\ \emph {et~al.}(2005)\citenamefont
  {Diederich}, \citenamefont {D{\"o}ppner}, \citenamefont {Fennel},
  \citenamefont {Tiggesb{\"a}umker},\ and\ \citenamefont
  {Meiwes-Broer}}]{Diederich2005}%
  \BibitemOpen
  \bibfield  {author} {\bibinfo {author} {\bibfnamefont {T.}~\bibnamefont
  {Diederich}}, \bibinfo {author} {\bibfnamefont {T.}~\bibnamefont
  {D{\"o}ppner}}, \bibinfo {author} {\bibfnamefont {T.}~\bibnamefont {Fennel}},
  \bibinfo {author} {\bibfnamefont {J.}~\bibnamefont {Tiggesb{\"a}umker}}, \
  and\ \bibinfo {author} {\bibfnamefont {K.-H.}\ \bibnamefont {Meiwes-Broer}},\
  }\href@noop {} {\bibfield  {journal} {\bibinfo  {journal} {Phys. Rev. A}\
  }\textbf {\bibinfo {volume} {72}},\ \bibinfo {pages} {023203} (\bibinfo
  {year} {2005})}\BibitemShut {NoStop}%
\bibitem [{\citenamefont {Eland}(1991)}]{Eland1991}%
  \BibitemOpen
  \bibfield  {author} {\bibinfo {author} {\bibfnamefont {J.}~\bibnamefont
  {Eland}},\ }\href@noop {} {\bibfield  {journal} {\bibinfo  {journal} {Laser.
  Chem.}\ }\textbf {\bibinfo {volume} {11}},\ \bibinfo {pages} {259} (\bibinfo
  {year} {1991})}\BibitemShut {NoStop}%
\end{thebibliography}
\end{document}